\documentclass [12pt]{article}
\usepackage{pslatex}
\usepackage{graphics}
\usepackage{amsfonts}
\usepackage{amsmath}
\usepackage[table]{xcolor}
\usepackage{hyperref}

\begin{document}

\title{The Structure of Scenarios}
\author{Guido Fioretti\\University of Bologna\\Management Department}
\maketitle

\begin{abstract}
Scenarios elicit possibilities that may be ignored otherwise, as well as causal relations between them. Even when too little information is available to assess reliable probabilities, the structure of linkages between evoked alternatives and perceived consequences can be analyzed by highlighting shared consequences of different alternatives or, conversely, diverging consequences of apparently similar alternatives. While according to current practice this structure is analyzed by exploring four possibilities obtained by crossing two macro-features, I illustrate the wider possibilities enabled by hypergraph analysis. An application is discussed.
\end{abstract}

\bigskip



\newpage

\section{Introduction}   \label{sec:introduction}

While probabilistic uncertainty refers to a given set of possibilities, decision-making may be plagued by the fear that the possibility set is not exhaustive, i.e., that something may happen, that one is not even able to conceive. Scenario planning can be regarded as a practice that attempts to restrict these ``unknown unknowns''  \cite{rumsfeld-11}  by engaging business strategists in extensive discussion and intentional search for non-obvious possibilities that might upset the received  wisdom \cite{barbierimasini-medinavasquez-00TFSC} \cite{chermack-vandermerwe-03F} \cite{roxburgh-09MKQ} \cite{bodwell-chermack-10TFSC} \cite{vervoort-bendor-kelliher-strick-helfgott-15F} \cite{feduzi-faulkner-runde-cabantous-loch-20AMR}. 

However,  a fairly clear divide separates those methods that aim at systematically collecting as much information as possible from as many sources as possible, from those that by design focus on outliers and unconventional perspectives \cite{amer-daim-jetter-13F} \cite{cordovapozo-rouwette-23F}. On the one hand, the systematic approach monitors as many points of view as possible, running the risk of neglecting or downplaying the one single correct outlier  \cite{eriksson-hallding-skanberg-22F}. On the other hand the intuitive approach focuses on outliers, running the risk of falling prey of points of view that are simply peculiar \cite{erdmann-sichel-yeung-15MKQ}.

The systematic approach is keen to assess probabilities; therefore, it ultimately reduces scenarios evaluation to computing expected values. By contrast, the intuitive approach refrains from using probabilities in order not to dilute idiosyncratic intuitions \cite{wilson-00TFSC} \cite{goodwin-wright-01JMS} \cite{wright-goodwin-09IJF} \cite{ramirez-selin-14F}. One point of view that reconciles these approaches is that the first one is appropriate if reliable probabilities can be measured on sufficiently large samples, whereas the second one is in order whenever subjective probabilities rest on too little empirical evidence to represent anything beyond personal opinions and educated guess --- the so-called \emph{known unknowns}  \cite{rumsfeld-11}.

Both approaches must  provide top-managers with a just handful of scenarios. Human limitations on short-term menory \cite{cowan-00BBS} suggest a number of scenarios between two and six,  most often four  \cite{amer-daim-jetter-13F}. However, the systematic approach has developed formal techniques (\emph{morphological analysis}) that condense scenarios to manageable numbers by evaluating their diversity and their vulnerability to the proposed policies \cite{carlsen-lempert-wikmansvahn-schweitzer-16EMS} \cite{lord-helfgott-vervoort-16F}.  By contrast,   intuitive approaches typically select two  critical dimensions along which  four scenarios are identified. This state of affairs is unsatisfactory  because nothing ensures that those two dimensions are coupled to relevant differences among strategic decisions.

Henceforth, I shall illustrate a purely structural representation technique that can be applied to the intuitive approach because it does not require any probability assessment. It simply analyzes the structure of the possibilities that connect the available alternatives to  possible consequences, highlighting classes of consequences that can be reached by apparently different alternatives while separating those alternatives that eventually lead to diverging consequences.

The rest of this paper is organized as follows. First,  \S~\ref{sec:scenario-structure} illustrates structural measures of scenarios. Subsequently, \S~\ref{sec:application} applies these measures to an empirical example. Finally, \S~\ref{sec:conclusions} concludes.

\section{Scenarios as Hypergraphs}  \label{sec:scenario-structure}

\begin{figure}
\center
\fbox{\resizebox*{0.7\textwidth}{!}{\includegraphics{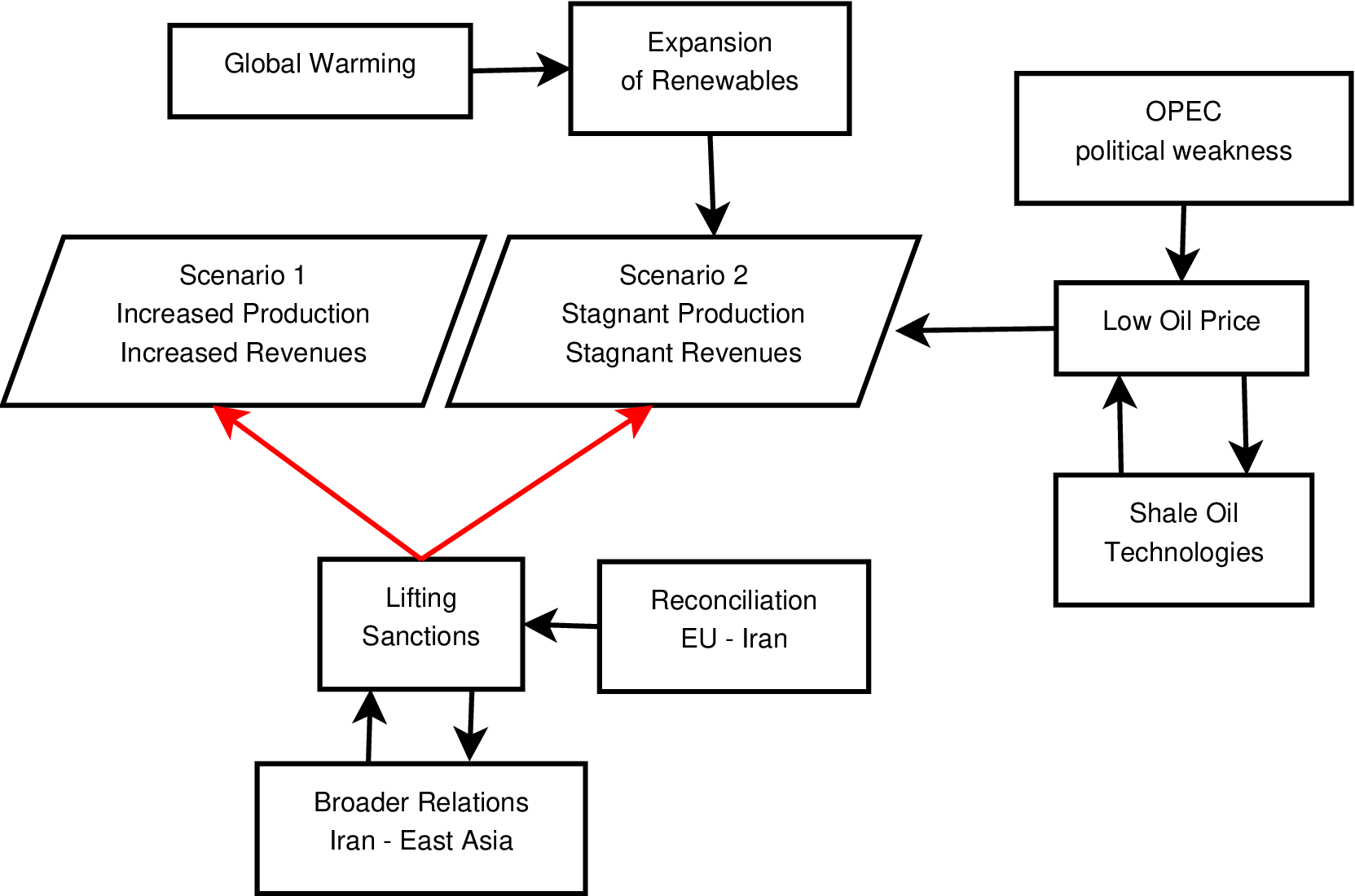}}}
\caption{One and the same action --- lifting sanctions to Iran --- may lead to quite different outcomes depending on many other factors, such as  growing availability of renewable energy sources or shale oil. One-to-many causal relations are highlighted in red. Loosely inspired by \cite{alipour-hafezi-amer-akhavan-17E}.} \label{fig:scenario}
\end{figure}

Scenarios can be seen as networks of concepts linked to one another by causal relations. Thus,  they are isomorphic to cognitive maps \cite{goodier-austin-soetanto-dainty-10F} \cite{amer-jetter-daim-11IJESM} \cite{jetter-schweinfort-11F} \cite{jetter-kok-14F} \cite{alipour-hafezi-amer-akhavan-17E}. For instance, Figure~(\ref{fig:scenario}) illustrates a possible cognitive map for the effects  of uplifting sanctions on Iran oil revenues, depending on the role played by renewable energy sources as well as shale oil (loosely inspired by \cite{alipour-hafezi-amer-akhavan-17E}). This map highlights that lifting sanctions to Iran may make this country increase its oil production, but that depending on circumstances stagnant oil production is equally possible. Notably, the value of this scenario exercise is  in highliting non-obvious outcomes  which may not be apparent at first sight.

The concepts that appear in a cognitive map may activate one another with differential strength, in which case one speaks of a fuzzy cognitive map. Fuzzy cognitive maps are widely employed in scenario planning \cite{jetter-schweinfort-11F} \cite{papageorgiou-13IEEE-TFS} \cite{jetter-kok-14F} but, since assessing strengths runs into the same difficulties as assessing probabilities, I shall focus on unweighted  maps  henceforth.

The causal linkages of cognitive maps can be passed through up to obtaining a Galois connection between two sets \cite{jetter-kok-14F} that represent   evoked alternatives (\textit{EA})  and perceived consequences (\textit{PC}), respectively  \cite{march-simon-58}. For instance, the scenario of Figure~(\ref{fig:scenario}) would be reduced to one evoked alternative (lifting sanctions) and two perceived consequences (either increasing or stagnant oil production, respectively).

\begin{figure}
\center
\fbox{\resizebox*{0.9\textwidth}{!}{\includegraphics{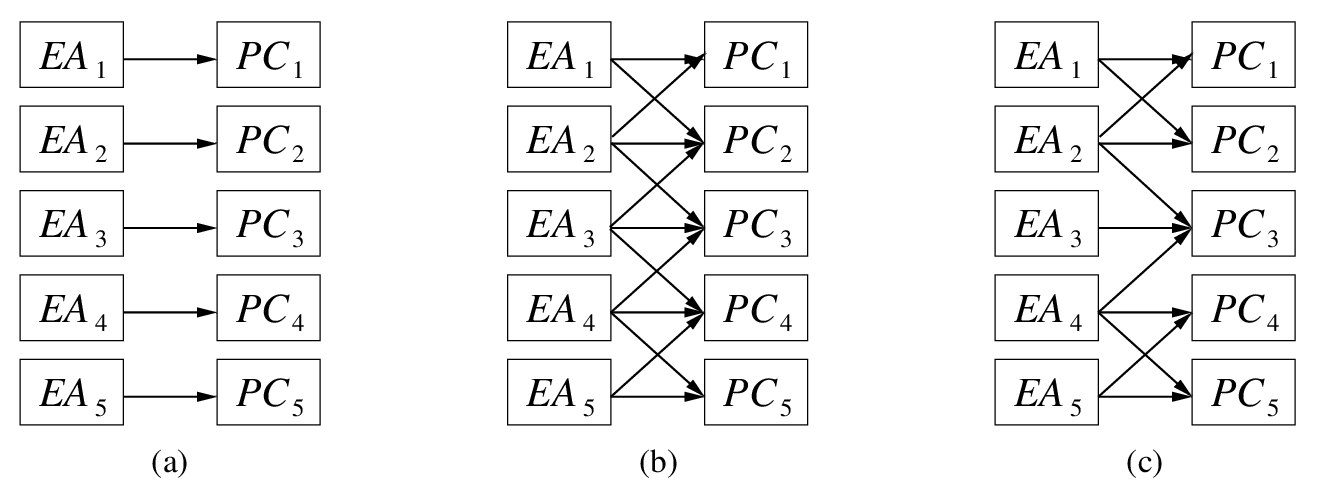}}}
\caption{Three stylized cognitive maps linking evoked alternatives (the $EA$s) to perceived consequences (the $PC$s). In (a), a one-to-one mapping with  $\mathcal{C}(\mathcal{K}) = 0$. In (b), the most confusing  one-to-many mapping where   $\mathcal{C}(\mathcal{K}) = 3$. In (c), a somewhat intermediate case where $\mathcal{C}(\mathcal{K}) = 2$.} \label{fig:one-to-many}
\end{figure}

Figure~(\ref{fig:one-to-many}) explains intuitively what structural reasonings are  possible. On the left  (a) the mapping between evoked alternatives and perceived consequences is one-to-one: This is the simple world where one perceives exactly which consequence follows from each of the evoked alternatives. Center (b), the extremely complex world where one perceives that any consequence can follow from each  of the  alternatives that are being evoked. On the right  (c) the somehow intermediate situation where, in spite of several one-to-many relations, one knows that certain evoked alternatives can only lead to a subset of the perceived consequences. The structural properties of (a), (b) and (c) can be measured, and compared to  one another by representing them as hypergraphs.~\footnote{A hypergraph is a generalization of a graph in which an edge (hyperedge) can connect any number of vertices. Alternatively, a  graph is a hypergraph whose hyperedges (edges) connect exactly two vertices.}

\begin{figure}
\center
\fbox{\resizebox*{0.8\textwidth}{!}{\includegraphics{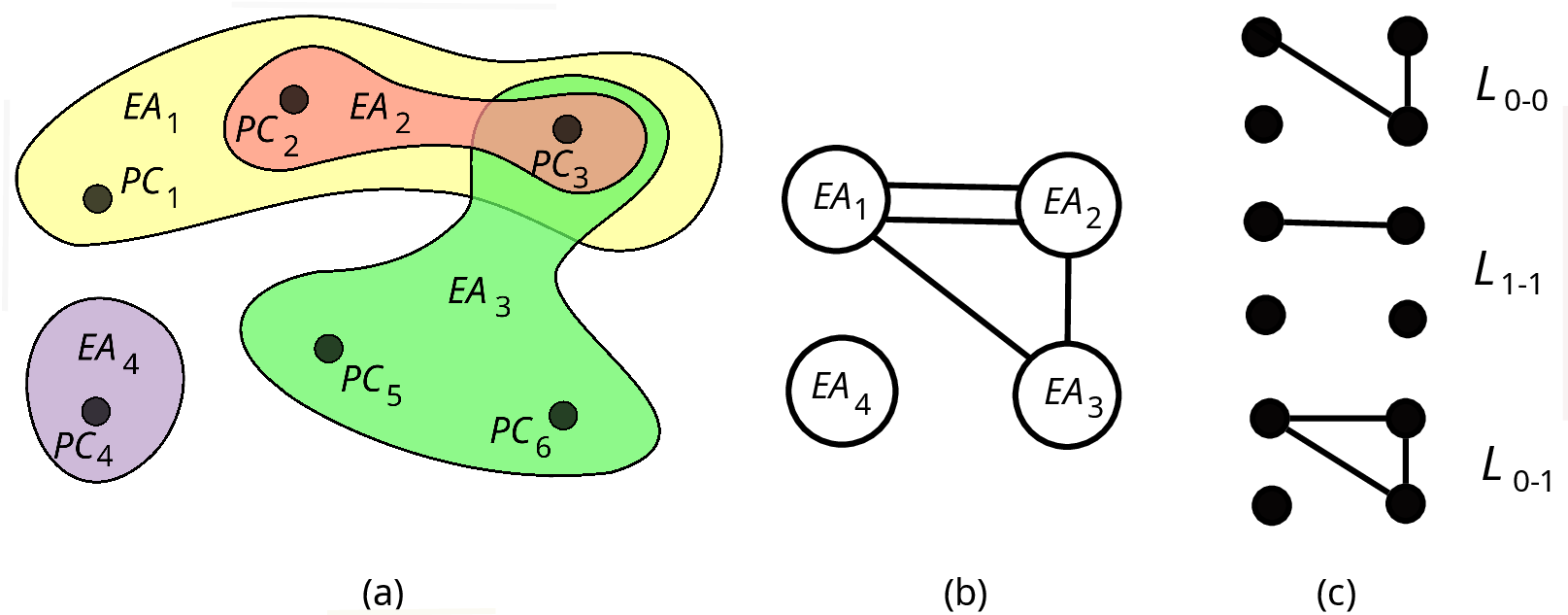}}}
\caption{Left (a), a hypergraph with $\{PC_j\}, \forall j = \{PC_1,  \ldots PC_6 \}$ and $\{EA_i\}, \forall i = \{EA_1, \ldots EA_4 \}$, with $EA_1 = \{PC_1,$ $PC_2,$ $PC_3 \}$ (yellow), $EA_2 = \{PC_2,$ $PC_3\}$ (red), $EA_3 = \{PC_3,$ $PC_5,$ $PC_6 \}$ (green) and $EA_4 = \{PC_4 \}$ (violet). Center (b), the multigraph obtained by condensing hyperedges into nodes. Right (c), the generalized line graphs $L_{0-0}$, $L_{1-1}$ and $L_{0-1}$, respectively. The generalized line graph  $L_{0-1}$ is eventually  called \emph{line graph}.} \label{fig:HG}
\end{figure}

Let us express the above mapping between the set of evoked alternatives $\{EA_i\}, \forall i$ and the set of perceived consequences $\{PC_j\}, \forall j$ by means of an \emph{undirected hypergraph} (henceforth, hypergraph) $\mathcal{H}$ where  $\{PC_j\}, \forall j$ is the set of vertices and the set of hyperedges is $\{EA_i\}, \forall i \subseteq \{PC_j\}, \forall j$. Figure~\ref{fig:HG}, section (a), illustrates a hypergraph where four hyperedges are highlighted in different colours.

Given a hypergraph $\mathcal{H}$, for any two hyperedges $EA_h$ and $EA_k$ such that $EA_h \cap EA_k \neq \emptyset$ let $p_{hk} = \| EA_h \cap EA_k  \|$ denote the dimension of their common face. Thus, $p=0$ means that they have one vertex in common (a face of dimension 0), $p=1$ means that they have two vertices in common (a face of dimension 1), and so on. Let $P$ denote the dimension of the highest-dimensional common face in $\mathcal{H}$.

Let us stipulate that  any two  hyperedges $EA_h$ and $EA_k$ are connected at level $q_{hk}$ if there exists a chain of hyperedges $\{EA_h, EA_u, EA_v, \ldots EA_w, EA_k\}$ such that $q_{hk} = p_{hu} = p_{uv} = \ldots p_{wk}$. In general, $q_{hk} \neq p_{hk}$. Let $Q \equiv P$ denote the highest dimension hyperedges are connected.

In general, $\forall q$ there exist $s_q$ disjoint classes of hyperedges pairwise connected at level $q$. The \emph{structure vector} $\mathbf{s}(\mathcal{H}) = [s_0 \; s_1, \ldots \; s_Q]$ subsumes the structural features of hypergraph $\mathcal{H}$. For instance, $s_1 = 2$ means that there exist two separate regions within which hyperedges have common faces that are made by  two vertices.

For any given hypergraph $\mathcal{H}$, a \emph{non-ordered simplicial complex} (henceforth, simplicial complex) $\mathcal{K}$ can be derived from  $\mathcal{H}$ by assuming that all faces of all hyperedges are themselves hyperedges. The hyperedges of a simplicial complex are called \emph{simplices}.~\footnote{The expression ``hypergraph'' derives from graph theory, whereas the expression ``simplicial complex'' derives from algebraic topology. I added the qualifications ``undirected'' and ``non-ordered'' because the literature is not always clear in this respect.} For instance, if the hypergraph depicted in Figure~\ref{fig:HG} is turned into a simplicial complex, then hyperedge $EA_1 = \{PC_1,$ $PC_2,$ $PC_3 \}$ entails hyperedges $\{PC_1,$ $PC_2\}$, $\{PC_1,$ $PC_3\}$, $\{PC_2,$ $PC_3\}$, $\{PC_1\}$, $\{PC_2\}$ and $\{PC_3\}$ as well, and all of them are called simplices. A structure vector $\mathbf{s}(\mathcal{K})$ can be similarly defined as above \cite{atkin-74} \cite{atkin-81} \cite{johnson-13}.

Usage of hypergraphs or simplicial complexes depends on context. Insofar scenario evaluation does not  depend on whether single consequences occur, or pair of consequences, or triplets and so on, simplicial complexes are the right object. 

The \emph{complexity} of $\mathcal{K}$ is an aggregate indicator of the extent to which the perceived consequences of different alternatives are separated from one another \cite{casti-89} \cite{fioretti-99ACS}:

\begin{equation} \label{eq:complexity} 
  \mathcal{C(\mathcal{K}}) \; = \; 
  \left\{
  \begin{array}{ll}
    0 & if \; all  \; 1:1 \\
  \sum_{q=0}^Q \; \frac{q+1}{\mathbf{s}_q(\mathcal{K})} & otherwise
  \end{array}
  \right.
\end{equation}
where complexity is zero and $\mathbf{s}_q \geq 1$. Note that, if a simplicial complex is connected, it is necessarily $\mathbf{s}_0 = 1$.

In the three cases illustrated in Figure~\ref{fig:one-to-many} it is $\mathbf{s}(\mathcal{K}_{\;\:a}) = [0 \; 0]$,  $\mathbf{s}(\mathcal{K}_{\;\:b}) = [1 \; 1]$ and $\mathbf{s}(\mathcal{K}_{\;\:c}) = [1 \; 2]$. Correspondingly, $\mathcal{C}(\mathcal{K}_{\;\:a}) = 0$,  $\mathcal{C}(\mathcal{K}_{\;\:b}) = 3$ and $\mathcal{C}(\mathcal{K}_{\;\:c}) = 2$. Thus, $\mathcal{C}(\mathcal{K})$ appears to be able to discriminate the case (c) as having lower complexity than (b). We shall see in \S~\ref{sec:application} that this property can be very useful because it allows to grasp that apparently messy scenarios do have a structure in fact.

Let $L_{p_* - p^*} (\mathcal{H})$ denote the \emph{generalized line graph} obtained from $\mathcal{H}$ by condensing hyperedges into nodes and by creating an edge between any two nodes $(h,k)$ if the corresponding hyperedges have a common face of dimension $p_{h,k} \in [p_*, p^*]$. In particular, a \emph{line graph}  $L (\mathcal{H})$ obtains when $p_* =0$ and $p^*=P$. Figure~\ref{fig:HG}, sections (b) and (c), illustrates the transformation of a hypergraph into generalized line graphs.

Similar concepts can be defined for simplicial complexes, except that if two simplices have a common face of dimension $p^*$, then they have common faces of any dimension $0 \leq p \leq p^*$. Line graphs on simplicial complexes will be denoted by $L_p^*(\mathcal{K})$ but, in general, $L_p^*(\mathcal{K}) \neq L_{0 - p^*}(\mathcal{H})$. Note that $L_p^*(\mathcal{K})$ connects all pairs $(EA_h, EA_k)$ such that $p_{hk} \geq p^*$. Henceforth, line graphs defined on simplicial complexes will be used to identify evoked alternatives that are connected to separate clusters of consequences.

\section{An Application}  \label{sec:application}

I illustrate the above concepts on a scenario analysis of alternative irrigation policies in Mexico \cite{lopez-morales-10}. This analysis implements a model of the Mexican economy using thirteen hydro-economic regions to capture the main features of the interface between water resources and economic activity.

A model of the Mexican economy generates a baseline scenario that reflects the current situation. This is compared to the consequences that can originate from four evoked alternatives:

\begin{itemize}

\item $EA_1$: Water withdrawal caps;  

\item $EA_2$: Positive water prices;  

\item $EA_3$: Water withdrawal caps + more efficient irrigation technologies;

\item $EA_4$: Positive water prices + more efficient irrigation technologies.

\end{itemize}

$EA_1$ consists of imposing caps on the amount of water that can be withdrawn from available sources for agricultural purposes. $EA_2$ aims at obtain similar outcomes by means of suitable prices that influence market  allocation in their turn. $EA_3$ and $EA_4$ are obtained  by adding more efficient irrigation techniques to $EA_1$ and $EA_2$, respectively.

\begin{figure}
\center
\fbox{\resizebox*{0.9\textwidth}{!}{\includegraphics{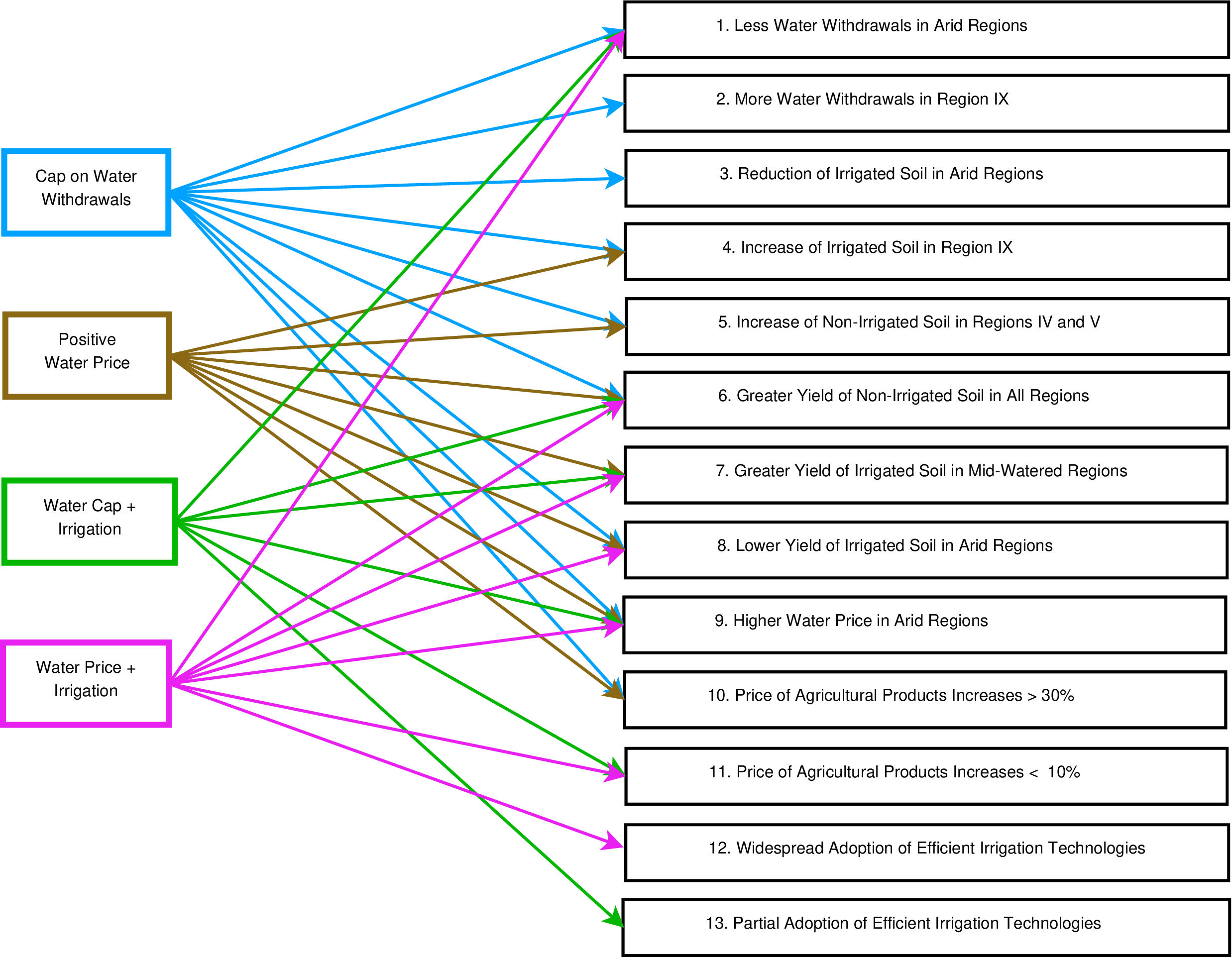}}}
\caption{Left, the four evoked alternatives. Right, the thirteen perceived consequences.} \label{fig:bipartite}
\end{figure}

Several consequences are perceived as possibly following from each of the alternatives that are being evoked. Figure~\ref{fig:bipartite} illustrates these causal relations.

\begin{figure}
\center
\fbox{\resizebox*{0.9\textwidth}{!}{\includegraphics{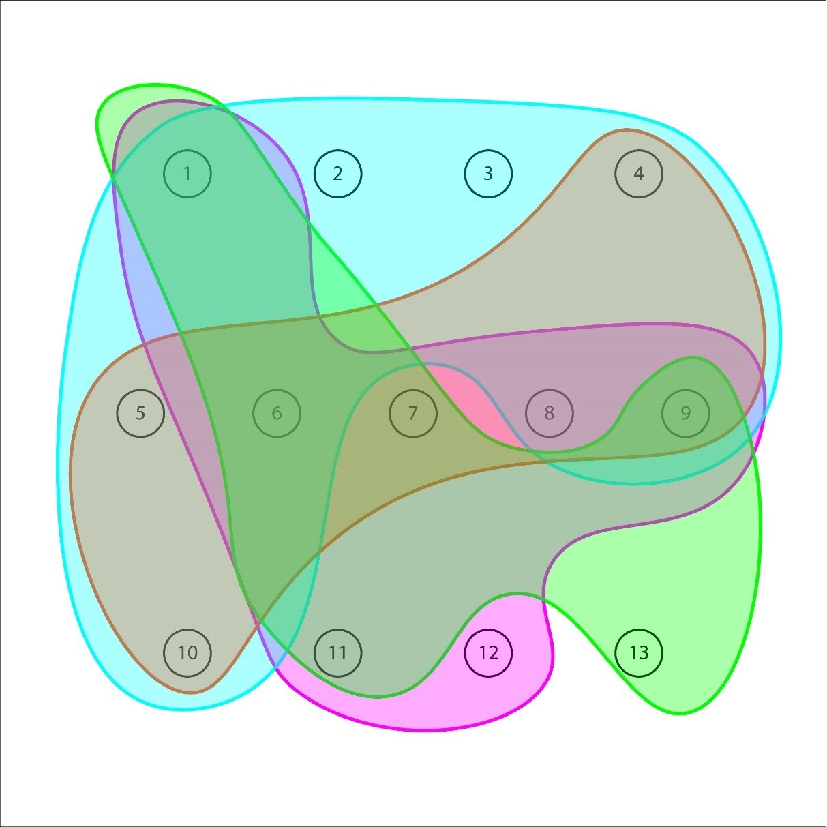}}}
\caption{The hypergraph corresponding to  Figure~\ref{fig:bipartite}. Evoked alternatives are drawn as colored hyperedges whose vertices are their perceived consequences. Evoked alternatives overlap insofar their share consequences.} \label{fig:hypergraph-Mexico}
\end{figure}

Let us represent this structure as a hypergraph whose hyperedges are the evoked alternatives $EA_1 = \{PC_1,$ $PC_2,$ $PC_3,$ $PC_4,$ $PC_5,$ $PC_6,$ $PC_8,$ $PC_9 \}$, $EA_2 = \{PC_4,$ $PC_5,$ $PC_6,$ $PC_7,$ $PC_8,$ $PC_9,$ $PC_{10}\}$, $EA_3 = \{PC_1,$ $PC_6,$ $PC_7,$ $PC_9,$ $PC_{11},$ $PC_{13}\}$, $EA_4 = \{PC_1,$ $PC_6,$ $PC_7,$ $PC_8,$ $PC_9,$ $PC_{11},$ $PC_{12}\}$ where vertices are the perceived consequences. Figure~\ref{fig:hypergraph-Mexico} is a graphical illustration where hyperedges merge their colors insofar they overlap.

Table~\ref{tab:overlaps} reports the dimension of the common face between any two hyperedges. From above, $EA_1$ and $EA_2$ have a common face of dimension $p=5$, $EA_1$ and $EA_3$ have a common face of dimension $p = 2$, and so forth.

\begin{table}    
\begin{center}
\begin{tabular}{c||c|c|c|c|} 

 \textit{Hyperedge} & $EA_1$ & $EA_2$ & $EA_3$ & $EA_4$ \\ \hline\hline
$EA_1$ & \cellcolor{gray} 8 & \cellcolor{yellow} 5 & 2 & 3  \\ \hline
$EA_2$ & \cellcolor{yellow} 5 & \cellcolor{gray} 6 & 2 & 3  \\ \hline
$EA_3$ & 2 & 2 & \cellcolor{gray} 5 & \cellcolor{yellow} 4  \\  \hline
$EA_4$ & 3 & 3 & \cellcolor{yellow} 4 & \cellcolor{gray} 5 \\ \hline
 \end{tabular}
\end{center}
\caption{The dimension of the common faces  between any two hyperedges. The elements on the diagonal represent the dimension of each hyperedge.} \label{tab:overlaps}
\end{table}

From Table~\ref{tab:overlaps} it is easy to reconstruct the classes of hyperedges of $\mathcal{H}$ connected at each level $q$, illustrated on the left side of Table~\ref{tab:classes}. Starting from $q=5$ and proceeding upwards one can compute the corresponding classes of simplices for $\mathcal{K}$ by assuming that each class at $q=5$ exists at $q=4$ as well, and so forth.

By counting the number of classes in $\mathcal{K}$ one obtains $\mathbf{s}(\mathcal{K}) = [1 \; 1 \; 1 \; 1 \; 2 \; 3]$ and, by applying eq.~\ref{eq:complexity}, $\mathcal{C}(\mathcal{K}) \equiv 14.5$. This is a benchmark to which variations of this scenario can be compared.

\begin{table}    
\begin{center}
\begin{tabular}{c|l|l} 
 $q$ & \multicolumn{1}{c}{$\mathcal{H}$} & \multicolumn{1}{c}{$\mathcal{K}$}  \\ \hline
0 & $\{EA_1\}$ $\{EA_2\}$ $\{EA_3\}$ $\{EA_4\}$ &  $\{EA_1, \; EA_2, \; EA_3, \: EA_4\}$ \\ 
1 & $\{EA_1\}$ $\{EA_2\}$ $\{EA_3\}$ $\{EA_4\}$ &  $\{EA_1, \; EA_2, \; EA_3, \: EA_4\}$ \\
2 & $\{EA_1, \; EA_2, \; EA_3\}$ $\{EA_4\}$ & $\{EA_1, \; EA_2, \; EA_3, \: EA_4\}$ \\ 
3 & $\{EA_1, \; EA_2, \; EA_4\}$ $\{EA_3\}$ & $\{EA_1, \; EA_2, \; EA_3, \: EA_4\}$ \\ 
4 & $\{EA_1\}$ $\{EA_2\}$ $\{EA_3, \: EA_4\}$ & $\{EA_1, \; EA_2\}$ $\{EA_3, \: EA_4\}$ \\
5 & $\{EA_1, \; EA_2\}$ $\{EA_3\}$ $\{EA_4\}$ & $\{EA_1, \; EA_2\}$ $\{EA_3\}$ $\{EA_4\}$ \\
 \end{tabular}
\end{center}
\caption{For each connection level $q$, the classes of hyperedges in $\mathcal{H}$ and the classes of simplices in $\mathcal{K}$.} \label{tab:classes}
\end{table}

When turning this hypergraph into a simplicial complex, the class of simplices connected at $q=5$   $\{EA_1$ $EA_2\}$ is connected at $q=4$ as well, and it does not overlap with the class  $\{EA_3$ $EA_4\}$. Thus, $s_4(\mathcal{K} = 2$. Since there are no further changes at lower connection levels except that all simplices are connected at $q=1$ and $q=0$, one obtain that  $\mathbf{s}(\mathcal{K}) = [1 \; 1 \; 1 \; 1 \; 2 \; 1]$. In this case, $\mathcal{C}(\mathcal{K}) \approx 18.5$.

\begin{figure}
\center
\fbox{\resizebox*{0.9\textwidth}{!}{\includegraphics{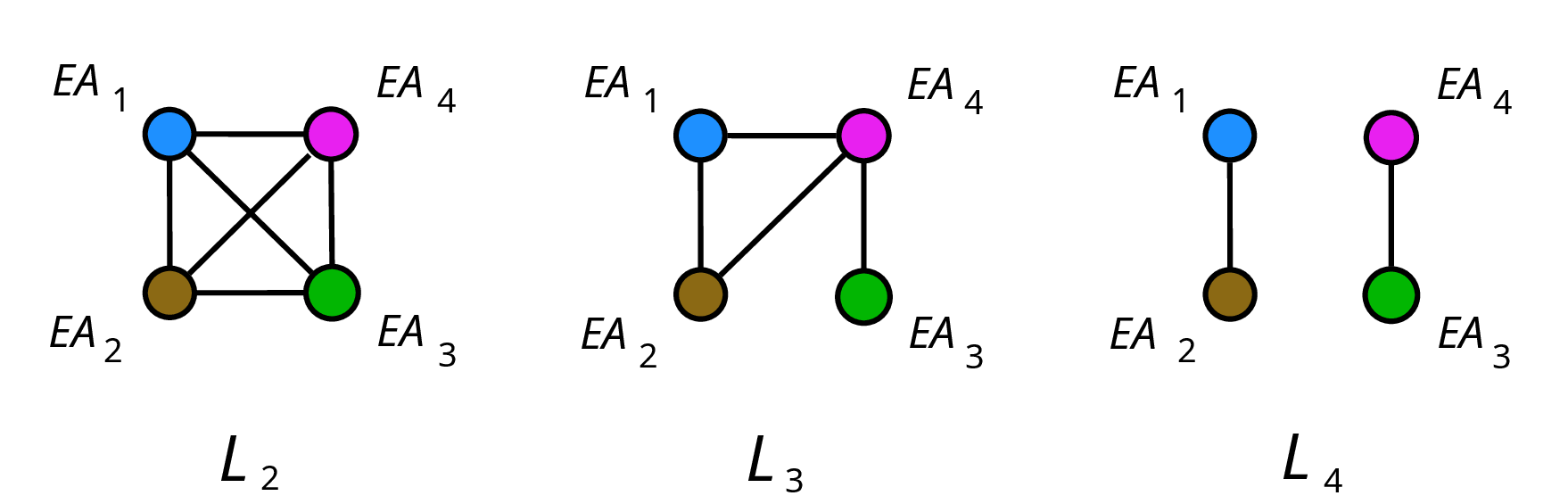}}}
\caption{Left to right, line graphs  $L_2(\mathcal{K})$,   $L_3(\mathcal{K})$ and $L_4(\mathcal{K})$, respectively.} \label{fig:linegraphs}
\end{figure}

Let us look at the details of the causal links from evoked alternatives to perceived consequences. Figure~\ref{fig:linegraphs} illustrates, left to right, the line graphs $L_2(\mathcal{K})$,   $L_3(\mathcal{K})$ and $L_4(\mathcal{K})$, respectively. Within this range, the structure of the simplicial complex changes from disorderly to structured.

While finding a disorderly structure up to $p^*=2$ implies that one can not rule out the possibility that any alternative generates any consequence, the appearance at $p^*=4$ of a structure separating $(EA_1, EA_2)$ from $(EA_3, EA_4)$ tells that the choice between these two pairs of alternatives is more robust than any choice within them.
In plain terms, the decision of introducing $(EA_3, EA_4)$  or not introducing  $(EA_1, EA_2)$ advanced irrigation techniques has clearer consequences than the choice between introducing water withdrawals caps and distorting water price.

This hypergraph was extremely simple, and similar conclusions could have been reached without the above analytical machinery. Let us complicate the picture.

\begin{figure}
\center
\fbox{\resizebox*{0.9\textwidth}{!}{\includegraphics{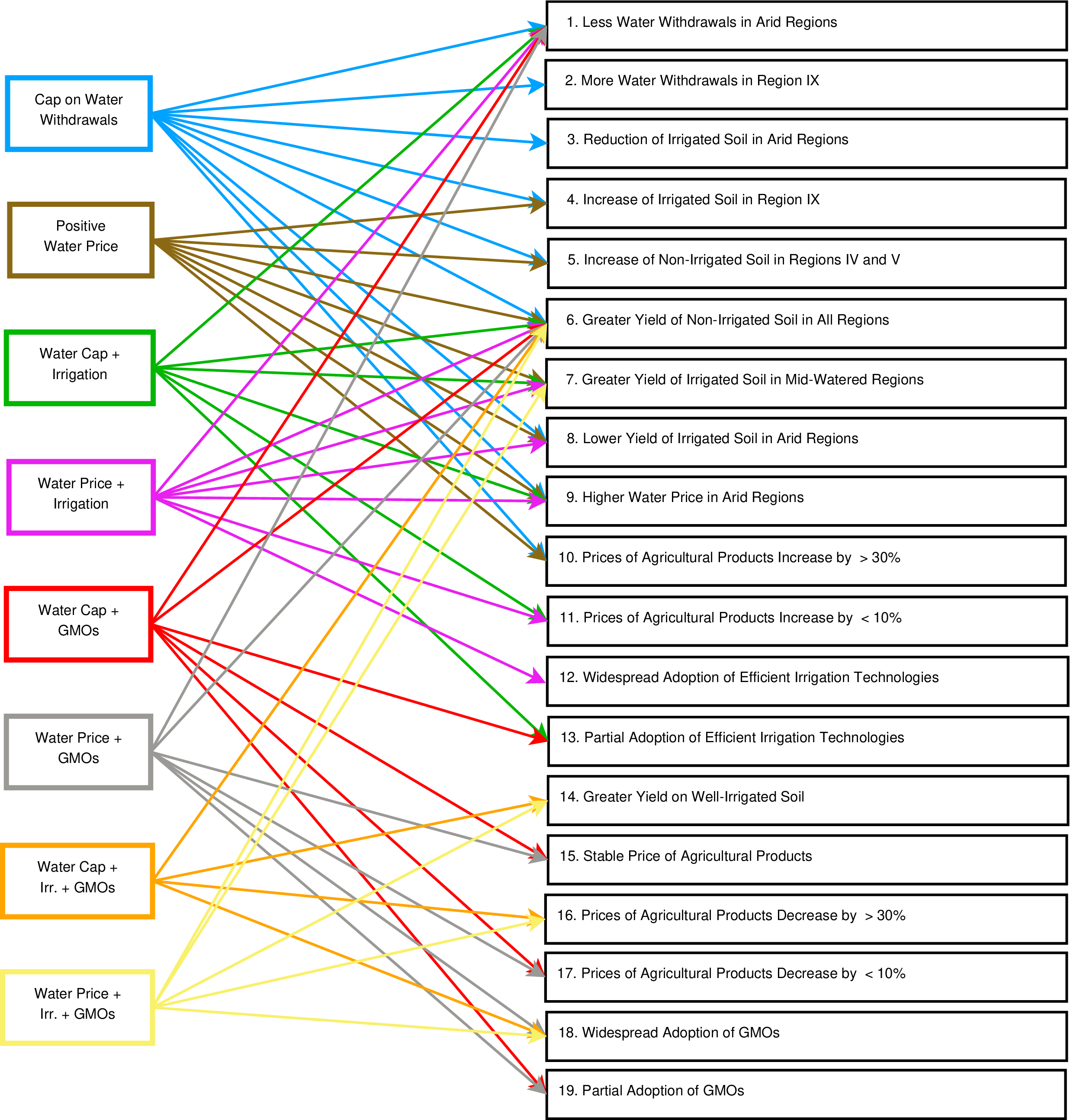}}}
\caption{Four new evoked alternatives have been added to the scheme of Figure~\ref{fig:bipartite}. GMOs can be either introduced alongside with either  water caps or  positive water price. Furthermore, both GMOs and advanced irrigation techniques can be combined with either  water caps, or  positive water price. Six new consequences are perceived in this scheme: (14) Greater yield on well-irrigated soil, (15) Stable price of agricultural products, (16) The price of agricultural products decreases by more than 30\%, (17)  The price of agricultural products decreases by less than 10\%, (18) Widespread adoption of GMOs, and (19) Partial adoption of GMOs.}  \label{fig:bipartite_GMOs}
\end{figure}

Let us suppose that the possibility of curbing water consumption by introducing genetically modified organisms (GMOs) is evoked. This would suffice to double the number of  alternatives from four to eight; correspondingly, several new consequences would be perceived. Figure~\ref{fig:bipartite_GMOs} illustrates one possible conceptual arrangement that adds new  alternatives and new  consequences to the scheme of Figure~\ref{fig:bipartite}.

A new hypergraph can be drawn, with twice as many hyperedges. Table~\ref{tab:overlaps_GMOs} reports the dimension of their common faces.

\begin{table}    
\begin{center}
\begin{tabular}{c||c|c|c|c|c|c|c|c|} 

 \textit{Hyperedge} & $EA_1$ & $EA_2$ & $EA_3$ & $EA_4$ & $EA_5$ & $EA_6$ & $EA_7$ & $EA_8$ \\ \hline\hline
$EA_1$ & \cellcolor{gray} 8 & \cellcolor{yellow} 5 & 2 & 3 & 1 & 1 & 0 & 0 \\ \hline
$EA_2$ & \cellcolor{yellow} 5 & \cellcolor{gray} 6 & 2 & 3 & 0 & 0 & 0 & 1  \\ \hline
$EA_3$ & 2 & 2 & \cellcolor{gray} 5 & \cellcolor{yellow} 4 & 2 & 1 & 0 & 1  \\  \hline
$EA_4$ & 3 & 3 & \cellcolor{yellow} 4 & \cellcolor{gray} 5 & 1 & 1 & 0 & 1 \\ \hline
$EA_5$ & 1 & 0 & 2 & 1 & \cellcolor{gray} 4 & \cellcolor{yellow} 4 & 0 & 0  \\ \hline
$EA_6$ & 1 & 0 & 1 & 1 & \cellcolor{yellow} 4 & \cellcolor{gray} 5 & 0 & 0  \\ \hline
$EA_7$ & 0 & 0 & 0 & 0 & 0 & 0 & \cellcolor{gray} 3 & 3  \\  \hline
$EA_8$ & 0 & 1 & 1 & 1 & 0 & 0 & 3 & \cellcolor{gray} 4 \\ \hline
 \end{tabular}
\end{center}
\caption{The dimension of the common face between any two hyperedges, including those added by considering GMOs.  The elements on the diagonal represent the dimension of each hyperedge.} \label{tab:overlaps_GMOs}
\end{table}

From Table~\ref{tab:overlaps_GMOs} the classes of connected hyperedges can be derived for each $q$. Table~\ref{tab:classes_GMOs} illustrates the corresponding classes for the simplicial complex.

\begin{table}    
\begin{center}
\begin{tabular}{c|l} 
 $q$ & \multicolumn{1}{c}{$\mathcal{K}$}  \\ \hline
0 & $\{EA_1, \; EA_2, \; EA_3, \; EA_4, \; EA_5, \; EA_6, \; EA_7, \; EA_8\}$  \\ 
1 & $\{EA_1, \; EA_2, \; EA_3, \; EA_4, \; EA_5, \; EA_6, \; EA_7, \; EA_8\}$  \\
2 & $\{EA_1, \; EA_2, \; EA_3, \; EA_4, \; EA_5, \; EA_6\}$ $\{EA_7, \; EA_8\}$  \\ 
3 & $\{EA_1, \; EA_2, \; EA_3, \; EA_4\}$ $\{EA_5, \; EA_6\}$ $\{EA_7, \; EA_8\}$  \\ 
4 & $\{EA_1, \; EA_2\}$ $\{EA_3, \: EA_4\}$ $\{EA_5, \; EA_6\}$ $\{EA_7\}$ $\{EA_8\}$  \\
5 & $\{EA_1, \; EA_2\}$ $\{EA_3\}$ $\{EA_4\}$ $\{EA_5\}$ $\{EA_6\}$ $\{EA_7\}$ $\{EA_8\}$  \\
 \end{tabular}
\end{center}
\caption{For each connection level $q$, the classes of simplices in $\mathcal{K}$ once the alternative of introducing GMOs has been evoked.} \label{tab:classes_GMOs}
\end{table}

The structure vector is $\mathbf{s}(\mathcal{K}) =$ $[1$ $1$ $2$ $2$ $5$ $7]$. By applying eq.~\ref{eq:complexity} one obtains $\mathcal{C}(\mathcal{K}) \approx 7.69$.

Notably, complexity decreased in spite of the fact that new alternatives have been evoked and new consequences are being perceived. This  signals that the additional alternatives and  consequences simplify the structure of possibilities, a feature that is not immediately evident by comparing Figure~\ref{fig:bipartite_GMOs} with Figure~\ref{fig:bipartite}. Let us look into the details of the line graphs.

\begin{figure}
\center
\fbox{\resizebox*{0.9\textwidth}{!}{\includegraphics{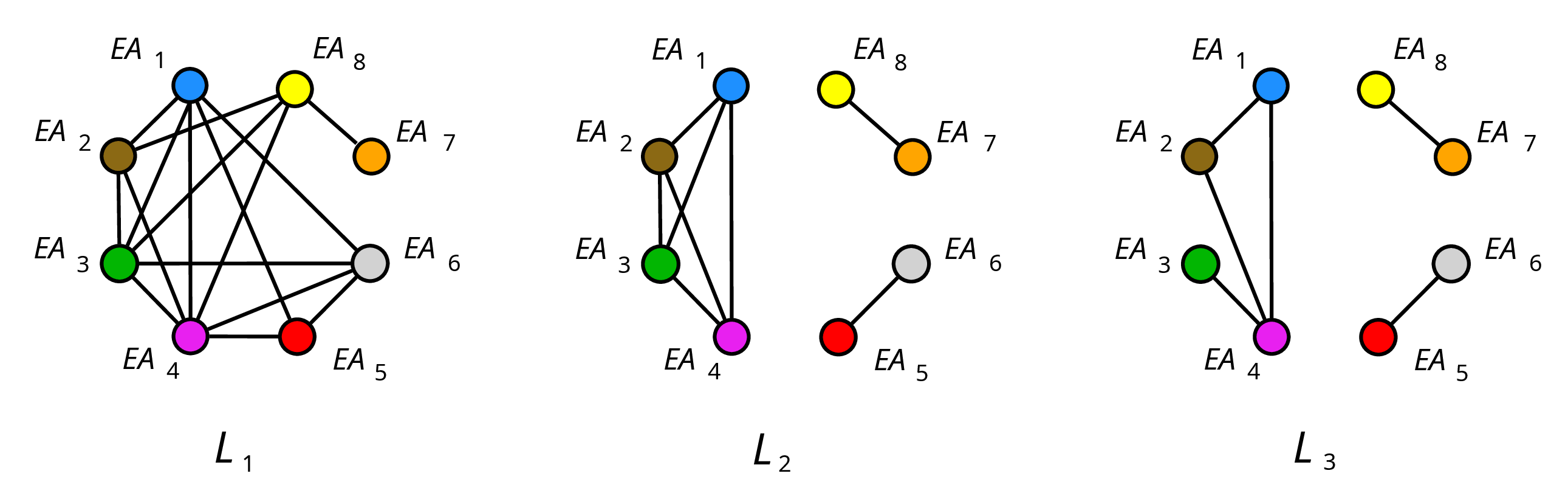}}}
\caption{Left to right, line graphs  $L_1(\mathcal{K})$,   $L_2(\mathcal{K})$ and $L_3(\mathcal{K})$, respectively.} \label{fig:linegraphs_GMO}
\end{figure}

Figure~\ref{fig:linegraphs_GMO} illustrates the line graphs of the simplicial complex at the transition from a disorderly state at $L_1(\mathcal{K})$ to a clear pattern that appears at  $L_2(\mathcal{K})$ and is roughly confirmed at $L_3(\mathcal{K})$. Several features of this pattern are neither obvious, nor easily discernible from Figure~\ref{fig:bipartite}.

In Figure~\ref{fig:linegraphs_GMO}, the two pairs of alternatives that introduce GMOs, either alone $(AE_5, AE_6)$ or in conjunction with advanced irrigation techniques $(AE_7, AE_8)$ are quite isolated from one another as well as from the more traditional alternatives that had been previously evoked. Within each pair of alternatives there is little differentiation between capping water withdrawals and distorting water price, and this is in line with the original set of scenarios; however, it is quite remarkable that coupling GMOs with advanced irrigation techniques generates consequences that are --- to a large extent --- qualitatively different from those obtained by GMOs alone as well as from advanced irrigation techniques alone.

This extension of the basic scenarios is fictional. Thus, its value does not reside in the conclusions that it yields but rather in showing that simplicial complex analysis is able to yield insights that remain hidden to visual inspection of  Figure~\ref{fig:linegraphs_GMO}.

According to current practice, such complex maps are simplified by selecting two dimensions with no explicit criterion. One may focus on irrigation techniques and GMOs, considering four quadrants: (a) Advanced irrigation only; (b) GMOs only; (c) Advanced irrigation + GMOs; (d) Neither advanced irrigation, nor GMOs, and the analysis would highlight many interesting features of the mapping between evoked alternatives and perceived consequences. Or, the four quadrants could be: ($\alpha$) Only withdrawals caps; ($\beta$) Only price distortion; ($\gamma$) Both withdrawal caps and price distortion; ($\delta$) Neither withdrawal caps, nor price distortion. Perhaps this is the focus that economic theory would suggest, but it would hide the most prominent features of the mapping between evoked alternatives and perceived consequences. Most importantly, there does not exist any criterion to focus on either $\{a,$ $b,$ $c,$ $d\}$ or $\{\alpha,$ $\beta,$ $\gamma,$ $\delta\}$. With $q-$analysis of simplicial complexes, the most prominent structural features of the mapping between evoked alternatives and perceived consequences are highlighted.

\section{Conclusions}    \label{sec:conclusions}

Cognitive maps  have been imported into scenario planning from artificial intelligence applications where they were eventually employed as a means to forecast the most likely future \cite{jetter-kok-14F}. This may have obscured the most important aim of futures studies, namely, highlighting non-obvious and yet possible future states.

The technique that I have illustrated is far from being mechanistic. If anything, it may be criticised for not providing clear-cut answers but rather a rough guide to the structure of causal relations that underlie scenario analysis.

Several investigations have highlighted that most companies use scenarios rather informally, and that this technique is most successful when managers are involved in their development from the very beginning \cite{linneman-klein-79LRP}. There are concerns over declining usage of scenarios, eventually ascribed to the distance of scenarios from the concrete decisions that companies are called to make  \cite{millet-03SL}. In short, while many scenarios are being designed to suggest visions, many companies  seek help with their short-term decisions \cite{courtney-03SL}.

There might be some myopia in underestimating the value of a vision, but if help with decision-making is asked, experts should possibly express visions in terms of the consequences that specific choices might generate. A technique does not make for a trend, but if a trend towards greater closeness of scenarios to day-to-day decisions, then hypergraphs might have a place in it.

\clearpage
\bibliographystyle{plain}
\bibliography{references}
\end{document}